\documentclass[prb,
twocolumn,
superscriptaddress,showpacs,amsmath,amssymb]{revtex4}
\usepackage{amsfonts}
\usepackage{bm}
\usepackage{verbatim}

\usepackage{graphicx}

\begin{document}

\title{Superfluids in rotation:  
Landau-Lifshitz vortex sheets vs Onsager-Feynman vortices}

\author{G.E. Volovik}

\affiliation{Low Temperature Laboratory, Aalto University, P.O. Box 15100, FI-00076 AALTO, Finland}

\affiliation{ L.~D.~Landau Institute for
Theoretical Physics, 117940 Moscow, Russia}

\date{\today}

\begin{abstract}
{ 
The paper by Landau and Lifshitz on vortex sheets in rotating superfluid appeared in 1955 almost at the same time when Feynman published his paper on quantized vortices in superfluid $^4$He. For a long time this paper has been considered as an error. But 40 years later the vortex sheets have been detected in chiral superfluid $^3$He-A in the rotating cryostat constructed in the Olli Lounasmaa Low Temperature Laboratory (Otaniemi, Finland).  The equation derived by Landau and Lifshits for the distance between the vortex sheets as a function of the angular velocity of rotation has  been experimentally confirmed, which is the triumph of the theory. We discuss different configurations of the vortex sheets observed and to be observed in superfluid $^3$He-A.
}
\end{abstract}

\maketitle

\section{Introduction: quantized vortices and vortex sheet}

Superfluid liquids were believed to obey the irrotational (potential) flow, $\nabla\times{\bf v}_{\rm s}=0$.
However, experiments by Andronikashvili and Osborne demonstrated formation of meniscus in superfluid $^4$He under rotation. This indicated that this superfluid rotates as a normal liquid, i.e. it participates in the solid body rotation, ${\bf v}_{\rm s}={\bf \Omega}\times{\bf r}$  and thus $\nabla\times{\bf v}_{\rm s}=2{\bf \Omega}$. To resolve this puzzle two scenarios were suggested in 1955:  by Feynman \cite{Feynman1955} and by Landau and Lifshitz \cite{paper1955}. 
In the Feynman approach
the solid body rotation on a macroscopic scale is imitated by he array of quantized vortices in Fig. \ref{4HeUnstable} ({\it top left}).  

\begin{figure}
\includegraphics[width=1.1\linewidth]{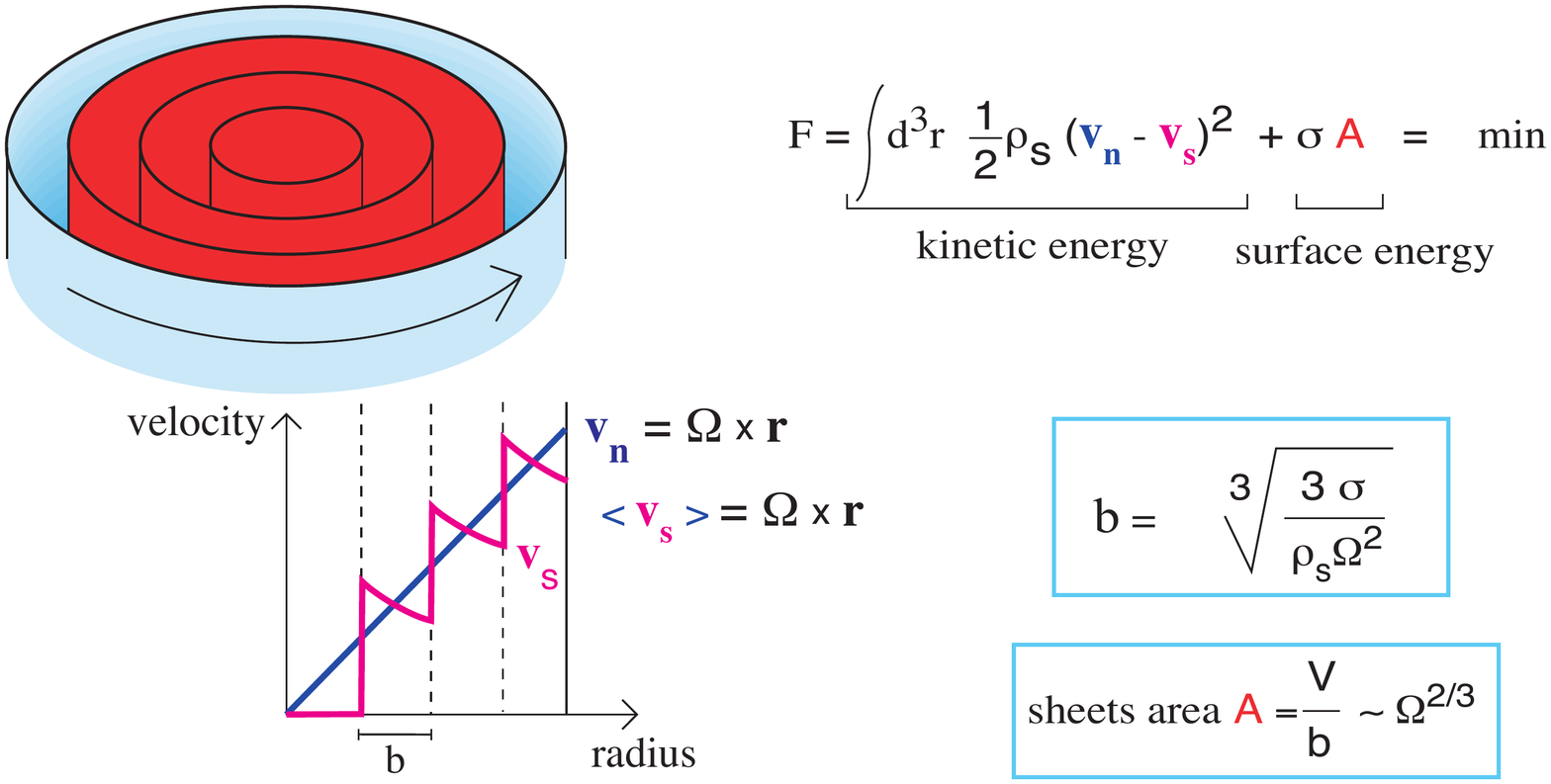}
\caption{
Vortex sheet scenario by Landau and  Lifshitz (1955).
Due to cylindrical vortex sheets 
the potential flow of superfluid component between the sheets
simulates the solid body rotation of liquid in average.
The distance $b$ between sheets as function of angular velocity $\Omega$ of rotation
is found by minimization of kiletic energy of flow and surface tension of the tangential discontinuity.
 }
 \label{LL}
\end{figure}

Landau and Lifshitz had another idea. They suggested that irrotational circulating flow was concentrated between the cylindrical vortex sheets (Fig. \ref{LL} {\it top right}). Historically,  vorticity concentrated in sheets was suggested by
Onsager in 1948 \cite{Onsager,London}  (see also the vortex sheet in the Onsager handwritten 1945 notes in Fig. 5 of Ref. \cite{Sreenivasan2006}).
In this arrangement the flow of superfluid component also corresponds to the solid body rotation on a
macroscopic scale  (Fig. \ref{LL} {\it bottom left}). 
By the minimization of energy  Landau and Lifshitz calculated the spacing $b$ between the  neighboring vortex sheets: 
\begin{equation}
b=\left(\frac{3\sigma}
{\rho_{s}\Omega^2}\right)^{1/3}
\,.
\label{b}
\end{equation}
It is determined by the balance of the kinetic energy of the superflow in rotating frame, ${1 \over 2} \rho_s ({\bf
v}_s-{\bf v}_n)^2$,  and the surface tension $\sigma$ of the sheets  (Fig. \ref{LL} {\it right}).

The surface tension of the tangential discontinuity in superfluid $^4$He has been estimated  by Ginzburg \cite{Ginzburg1955}, and the effective density of the rotating superfluids with vortex sheets has been calculated by I.M. Lifshitz and M.I. Kaganov. \cite{ILifshitz1955}

\begin{figure}
\includegraphics[width=1.0\linewidth]{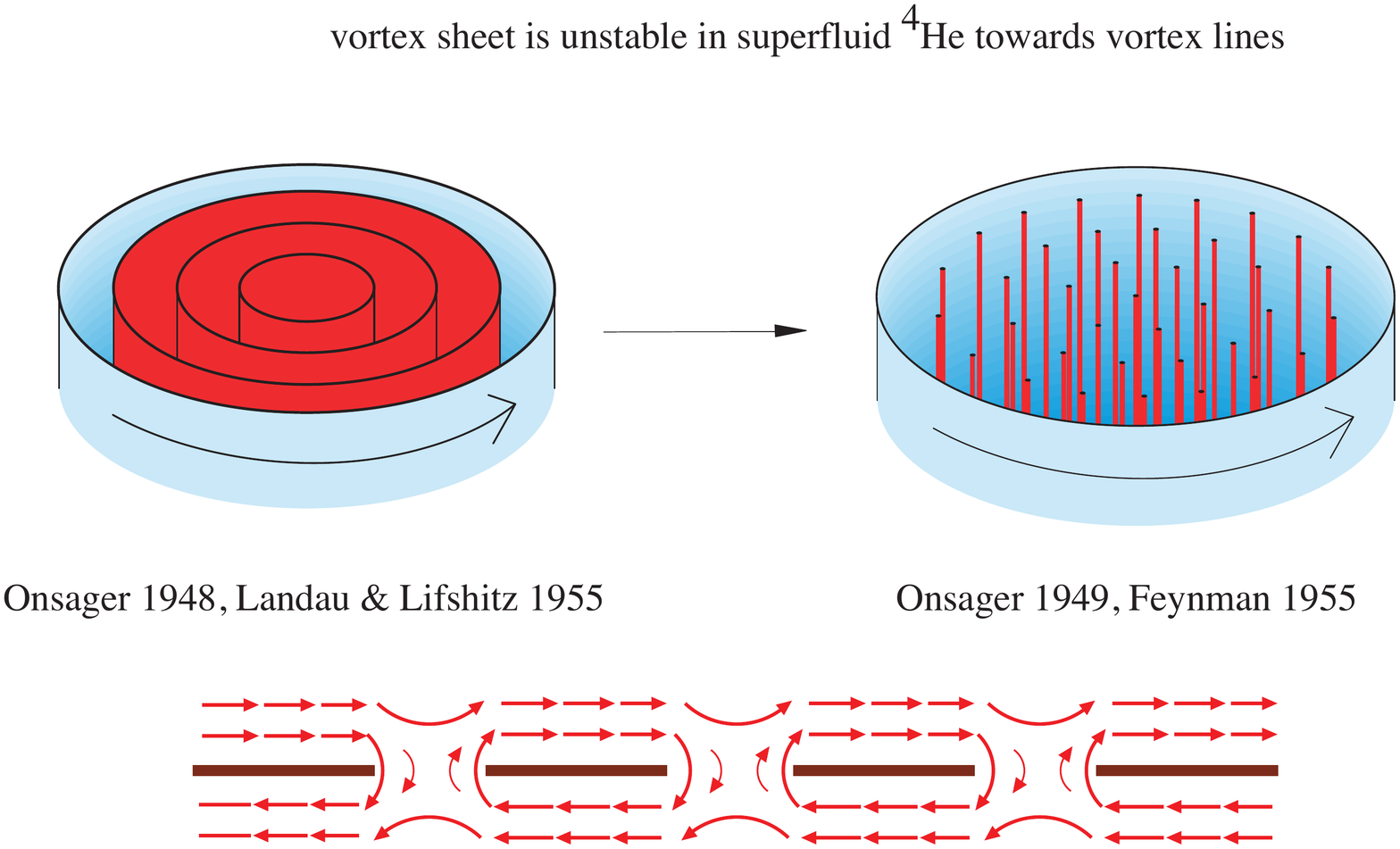}
\caption{
Vortex sheet is unstable in superfluid $^4$He towards vortex lines.
The tangential discontinuity is unstable towards break-up of the sheet into separate pieces ({\it bottom}), which form quantized vortex lines ({\it top right}). 
 }
 \label{4HeUnstable}
\end{figure}

\section{Vortex sheet instability in superfluid $^4$He}

It turned out, however,  that  in superfluid $^4$He the vortex sheet scenario does not work for several 
reasons. First, the tangential discontinuity is unstable towards break-up of the sheet into separate pieces (Fig. \ref{4HeUnstable} {\it bottom}), which finally transform to the quantized vortex lines in Fig. \ref{4HeUnstable} ({\it top right}). 

Then, there is the nucleation problem. According to Landau and  Lifshitz  the system of the equidistant layers should exist at sufficiently high velocity $\Omega$ of rotation. At low $\Omega$ a small cylindrical sheet appears first.  As we know now the objects are not easily created in the bulk liquid: they appear on the boundary and then propagate to the bulk. This would mean that with increasing angular velocity, the elementary cylindrical sheets start to penetrate into rotating vessel. Moreover, at high velocity the array of small sheets is energetically more favorable than the system of the coaxial sheets.  All this suggests, that the scenario of the coaxial vortex sheets may never occur.

Nevertheless,  though the Landau-Lifshitz  scenario was not applicable to the trivial superfluidity in liquid $^4$He, this idea turned out to be exactly to the point  for chiral superfluid $^3$He-A. 
First, the vortex sheet there is locally stable, being based on topological soliton. Second, the sheet is formed from the boundary of container. And finally,
 though the system of small cylinders and the system of quantized vortices are energetically more favorable, they are not formed in the process of the growth of the vortex sheet state.

The Landau-Lifshitz equation (\ref{b}) for distance between the vortex sheets as a function of $\Omega$ has  been experimentally confirmed in the NMR measurements, which was the triumph of the theory. The Bragg reflection from the
locally equidistant sheet planes has been also found \cite{Parts1994b}.

\begin{figure}
\includegraphics[width=1.0\linewidth]{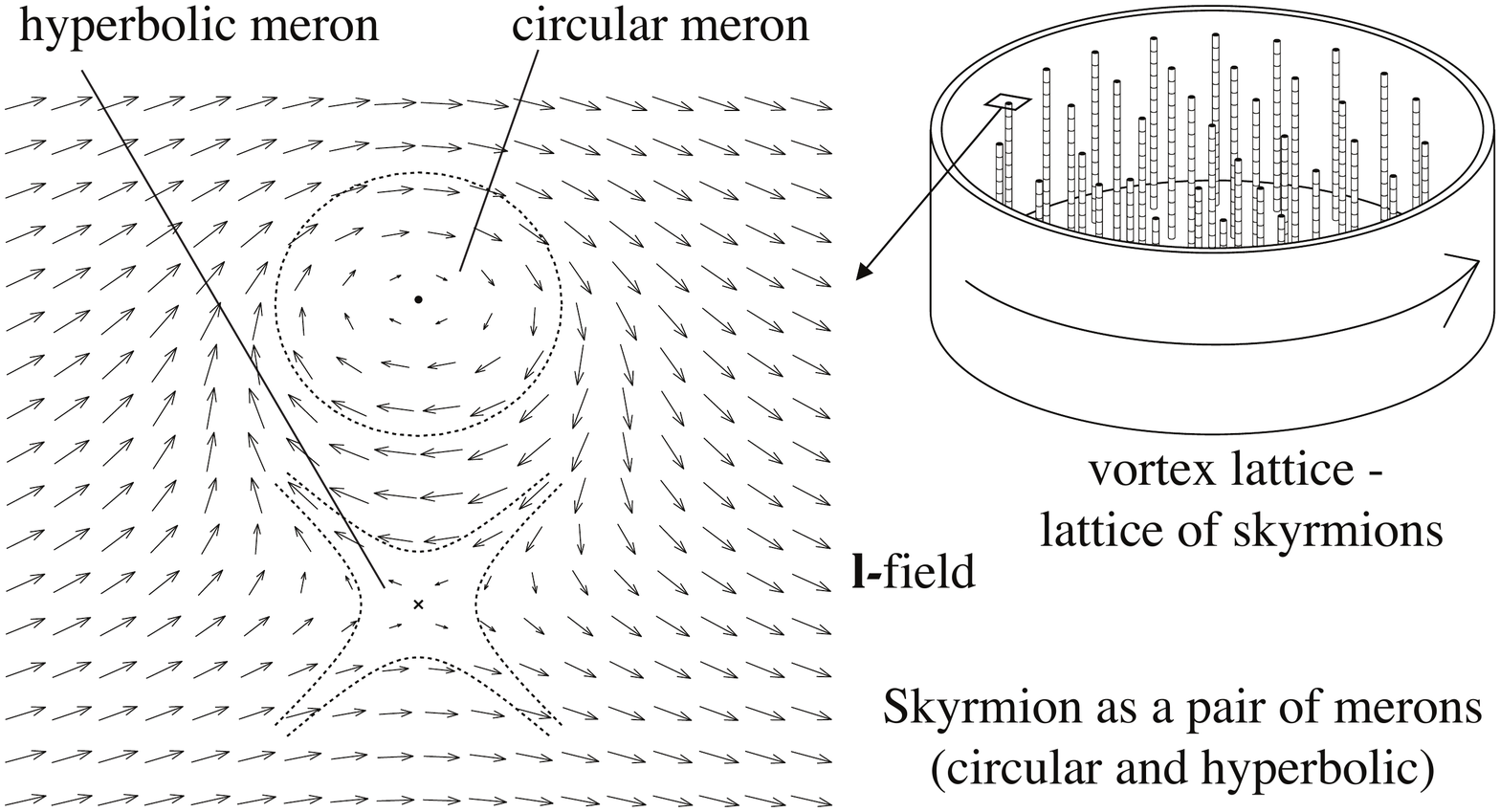}
\caption{Superfluid $^3$He-A is orbital ferromagnet with magnetization along the orbital 
angular momentum ${\hat{\bf l}}$ of Cooper pair, and spin antiferromagnetic (spin nematic) with anisotropy axis ${\hat{\bf d}}$. Typical vortices, which are created in $^3$He-A  under rotation, are continuous vortices with $4\pi$ winding of the condensate phase around the vortex (two quanta of circulation). The vorticity is generated by texture in the ferromagnetic 
field ${\hat{\bf l}}$, which is called the skyrmion. Skyrmion can be represented as the bound state of two merons, circular and hyperbolic. Each meron represents a vortex with a single quantum of circulation (the so-called Mermin-Ho vortex \cite{Mermin-Ho}).   In the circular vortex-meron $\hat{\bf l} \parallel \mathbf{\Omega}$ in the center, while in the hyperbolic vortex-meron $\hat{\bf l} \parallel - \mathbf{\Omega}$ in the center.
 }
 \label{Skyrmion-figure}
\end{figure}

\section{Continuous vorticity in chiral superfluid $^3$He-A:  
skyrmions and merons}

The chiral superfluid $^3$He-A is an orbital ferromagnet with magnetization along the orbital 
angular momentum ${\hat{\bf l}}$ of Cooper pair \cite{VollhardtWolfle1990}. Simultaneously  it is the spin antiferromagnetic (spin nematic) with anisotropy axis ${\hat{\bf d}}$, which allows us to investigate the properties of $^3$He-A using NMR.

The typical vortices, which appear in the chiral superfluid $^3$He-A under rotation, are continuous vortices -- textures without singularity in the order parameter field, Fig. \ref{Skyrmion-figure} ({\it left}). The structure of the  ${\hat{\bf l}}$-field in the continuous vortex  is similar to the structure of skyrmions -- the topologically twisted continuous field configurations in quantum field theory  \cite{Skyrme1962}.  The topological winding number of the skyrmion texture gives rise to the quantized circulation of superfluid velocity around the skyrmion  \cite{Mermin-Ho}. Skyrmion represents the  vortex with two quanta of circulation  ($N=2$), i.e. the phase of the order parameter changes by  $4\pi$, when circling around the vortex.

Skyrmion consists of 
two radicals, called merons, each with the circulation number $N=1$ (the $2\pi$-vortex). 
The isolated meron cannot exist in the bulk liquid
for topological reasons. Vortex-skyrmions have been experimentally  identified in $^3$He-A in NMR experiments 
\cite{Seppala1984}.
The lattice of skyrmions has been later discovered also in magnetic materials \cite{Muhlbauer2009}.

\begin{figure}
\includegraphics[width=1.0\linewidth]{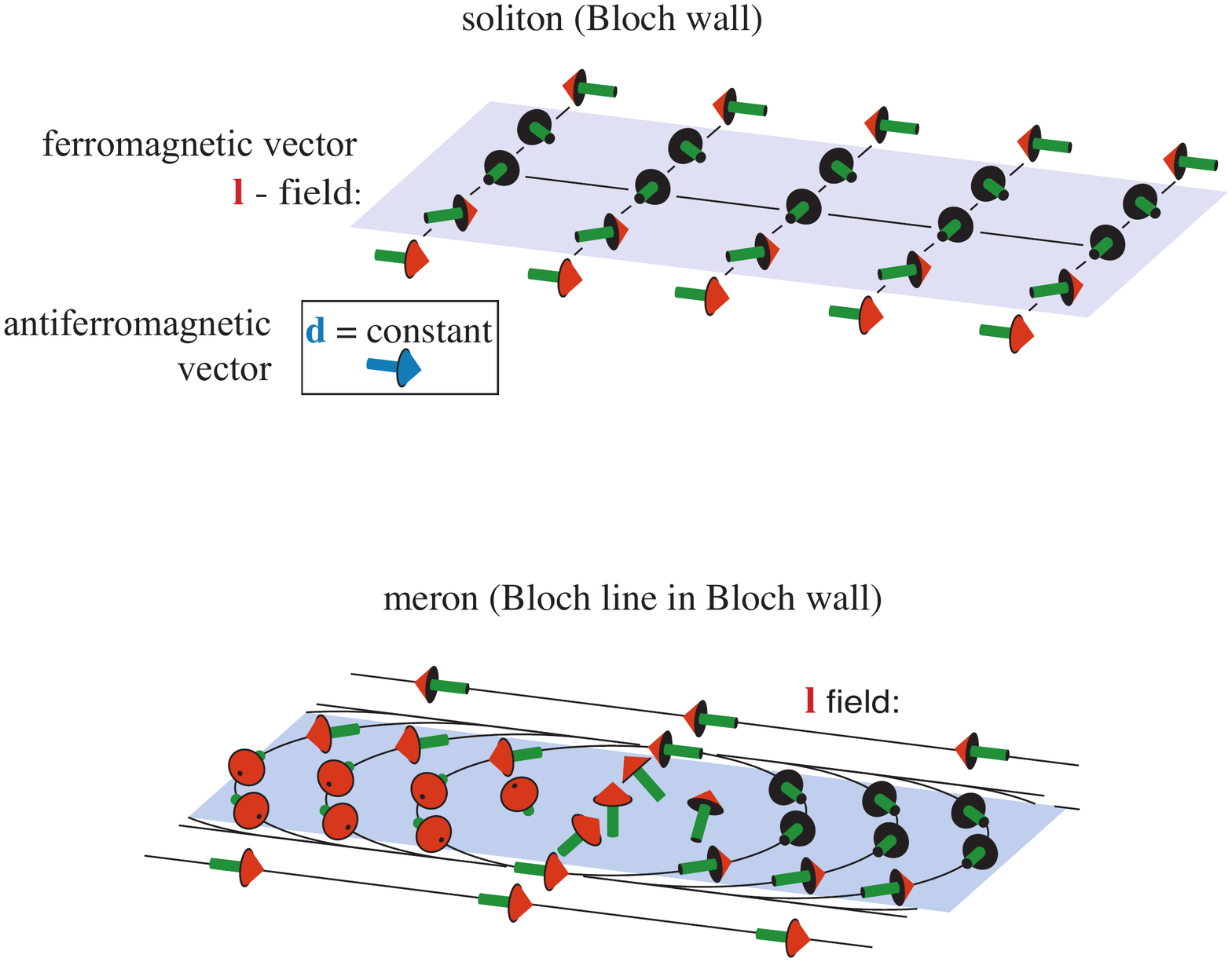}
\caption{
 {\it top}: Topological soliton in superfluid $^3$He-A is the analog of the Bloch or Neel wall in ferromagnets. It is characterized 
by the nontrivial element of $Z_2$ relative homotopy group. Figure demonstrates the  ${\hat{\bf l}}$-soliton, in which the ferromagnetic vector ${\hat{\bf l}}$ changes the direction to the opposite, while the antiferromagnetic vector ${\hat{\bf d}}$ remains constant.  As distinct from the domain wall, 
the soliton can terminate on a singular topological defect --  the vortex with $\pi$ winding number ($N=1/2$, the half-quantum vortex \cite{VolovikMineev1976,Volovik2003}), see Fig. \ref{Hole-figure}.
{\it bottom}: Bloch line within Bloch wall represents meron, which is the $2\pi$ vortex ($N=1$).
 }
 \label{Bloch-figure}
\end{figure}

\section{Solitons and merons}

The other possibility of existence of merons  is within the core of the
topological  soliton.  Topological $\hat {\bf l}$-soliton in superfluid $^3$He-A in Fig. \ref{Bloch-figure}
({\it top}) looks similar to the Bloch or Neel wall in ferromagnets. However, its topology is different: it is characteraized 
by the nontrivial element of $Z_2$ relative homotopy group.
Meron appears there in the  form of a kink in the $\hat {\bf l}$-soliton,  
Fig. \ref{Bloch-figure} ({\it bottom}).  The kink  separates two parts of the same soliton, which have
the opposite parities.  The solitonic meron looks similar to
the Bloch line within the Bloch wall  in ferromagnets \cite{Dedukh1985}, and it  
also cannot escape from the soliton (domain) wall.

For us it is important that merons carry the quantum of vorticity, $N=1$, and thus they serve  as the building blocks for construction of the vortex sheets, which are experimentally investigated in superfluid $^3$He-A \cite{Parts1994a,Parts1994b,Eltsov2000}.

\begin{figure}
\includegraphics[width=1.0\linewidth]{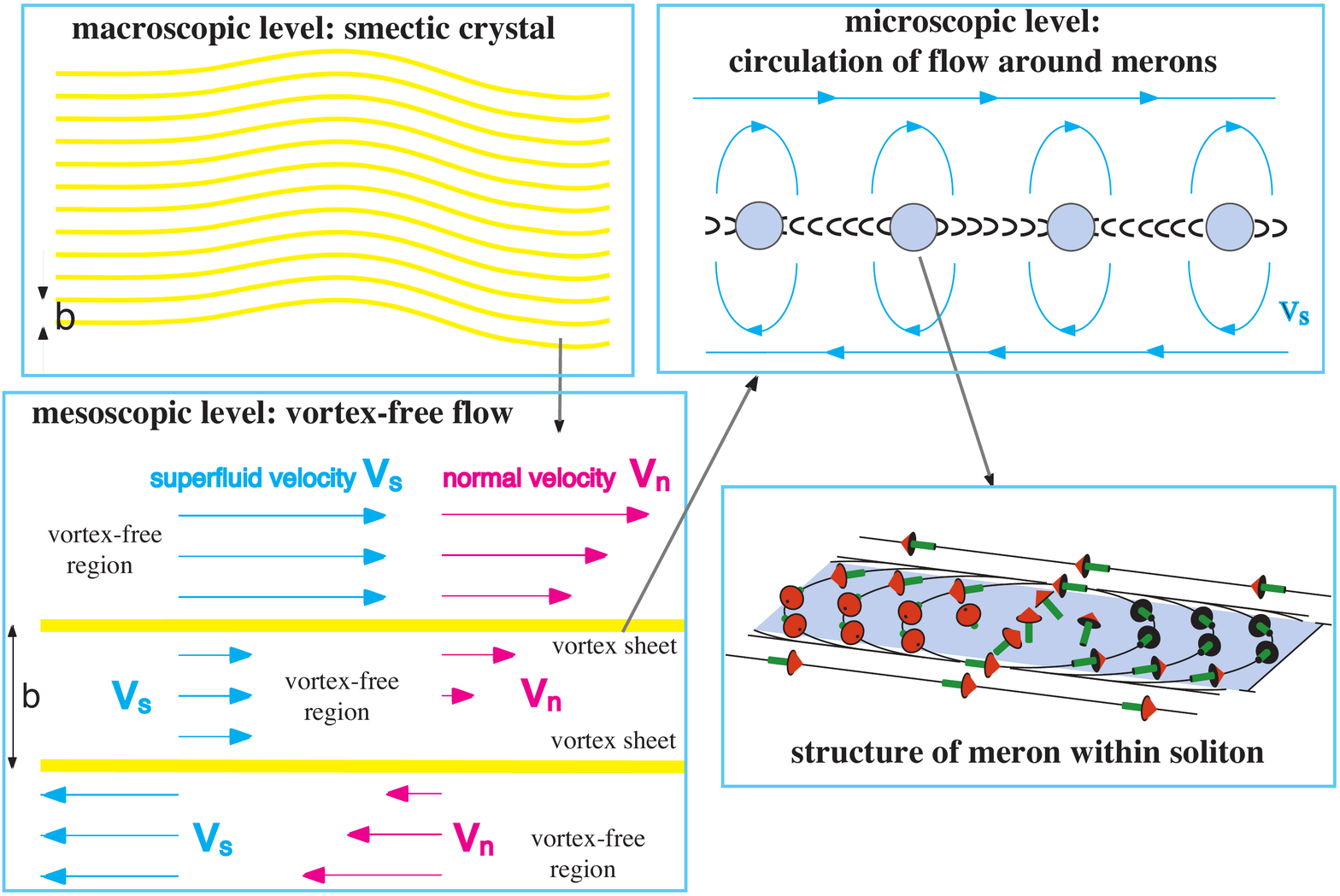}
\caption{Vortex sheet structure.
 ({\it top left}): On macroscopic scale the planes of the vortex sheets form  a local order of
the smectic liquid crystal.
 ({\it bottom left}): Flow velocity fields of superfluid and normal components between the sheet planes.
In these regions the superflow is vortex-free (superfluid velocity is constant), while the  velocity of the normal component ${\bf v}_{\rm n}=2\Omega y \hat{\bf x}$ with $\nabla\times {\bf v}_{\rm n}=2{\bf \Omega}$.
 ({\it top right}): Flow velocity field of superfluid component around merons (Mermin-Ho $N=1$ vortices).
({\it botton right}): Structure of  an individual circular meron.
 }
 \label{scales-figure}
\end{figure}

\section{Vortex sheet as a chain of merons}

When the topologically stable soliton accumulates  
vorticity in the form of merons  with the same circulation number $N=1$, it
forms
the vortex sheet, Fig.\ref{scales-figure}   ({\it top right}).  
 The vortex sheet has a hierarchy of length scales \cite{Heinila1995,Thuneberg1995}.  On a macroscopic scale the planes of the vortex sheets have a local order of the smectic liquid crystal,  Fig.\ref{scales-figure}   ({\it top left}).   The neighbouring planes of the vortex sheet are separated by
layers with the vortex-free superflow, Fig.\ref{scales-figure}   ({\it bottom left}). 

According to Landau-Lifshitz theory, the vortex sheets allow for the solid-body like rotation of the superfluid on a
macroscopic scale, i.e. with the average velocity field obeying equation 
$\left<{\bf v}_{\rm s}\right>= {\bf \Omega}\times{\bf r}$,    and thus 
$\left<\nabla\times {\bf v}_{\rm s}\right>=2{\bf \Omega}$.

\begin{figure}
\includegraphics[width=1.0\linewidth]{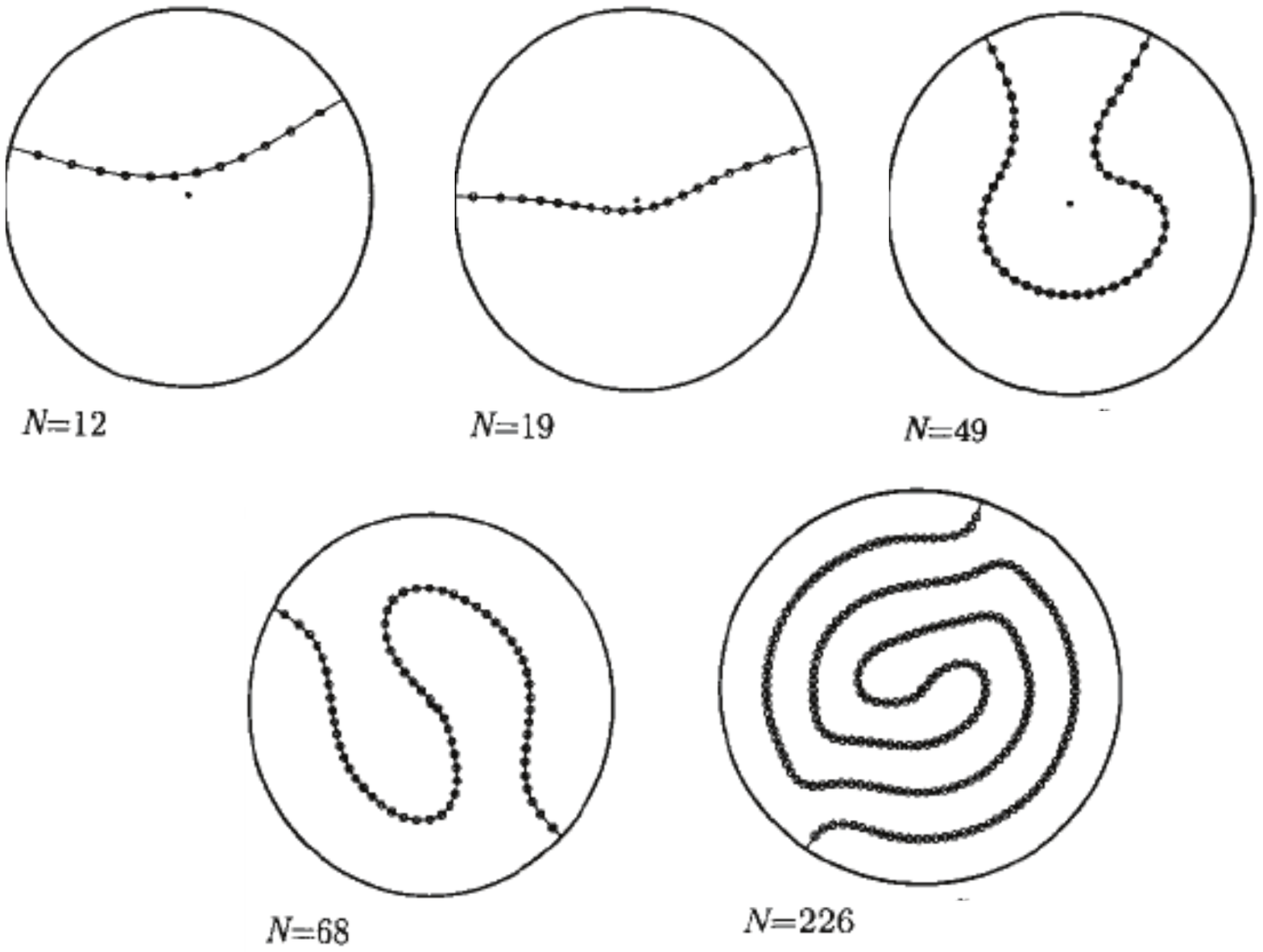}
\caption{
Numerical simulation of the growth of the vortex sheet with adiabatically  increasing angular velocity of rotation 
$\Omega$. $N$ is the number of merons in sheet. Vortex sheet  is formed in the rotating vessel, if the topological soliton is present in the cell. With increasing $\Omega$, merons
enter the soliton from  the side wall, and  the folded vortex sheet is formed.
For large $\Omega$ the configuration approaches that of the Landau-Lifshitz  cylindrical vortex sheets, but these sheets are connected. The dependence of the number of vortex-merons $N(\Omega)$ on angular velocity approaches the function $N_0(\Omega) =(2m/\hbar)\Omega R^2$, which does not depend on the surface tension of the soliton and corresponds to the equilibrium number of the isolated vortices in the rotating container.
In the range $12<N<19$ the vortex sheet has the form of the plane along the
diameter of the cylinder, see Fig. \ref{Planar-figure}. }
 \label{growth-figure}
\end{figure}


\section{Growth of the folded vortex sheet. Theory}

Numerical simulation of the growth of the vortex sheet \cite{Heinila1995} is shown in  Fig. \ref{growth-figure}. 
It follows from the experimental observations that once the vortex sheet
appears in the cell, the critical velocity of entering new 
merons into the cell is essentially  smaller than the critical velocity of the formation of skyrmions (about 6-7 times less). 
Merons have a small critical velocity for creation because the soliton is always in a close
contact with the side wall. The new merons are created at the edges of the
soliton at the wall and then easily enter the soliton. That is why the further  acceleration of the angular velocity leads to the growth of the soliton rather  than to the  creation of new skyrmions. In the final state of such adiabatic growth the skyrmions are absent, while the multiply  folded vortex sheet is formed, which locally simulates the coaxial sheets of 
Landau-Lifshitz, see  Fig. \ref{growth-figure} ({\it bottom right}). Note that in this figure the maximal number of merons was $N=226$, while in experiments $N$ can reach $10^3$, where the coaxial cylindrical structure 
of the folded vortex sheet is more pronounced.
Similar structure is suggested for the vortex sheet in the multi-component Bose condensates 
\cite{Simula2011}.

The number of merons in the vortex sheet depends on angular velocity $\Omega$ of the rotating vessel.
For large $\Omega$ the number of vortices-merons $N(\Omega)$ in the cell approaches the value $N_0(\Omega) =(2m/\hbar)\Omega R^2$, where $m$ is the mass of the $^3$He atom. This function $N_0(\Omega)$ does not depend on the surface tension of the soliton, it corresponds to the solid body rotation of the superfluid component, $\left<\nabla\times {\bf v}_{\rm s}\right>=2{\bf \Omega}$. 

\begin{figure}
\includegraphics[width=1.0\linewidth]{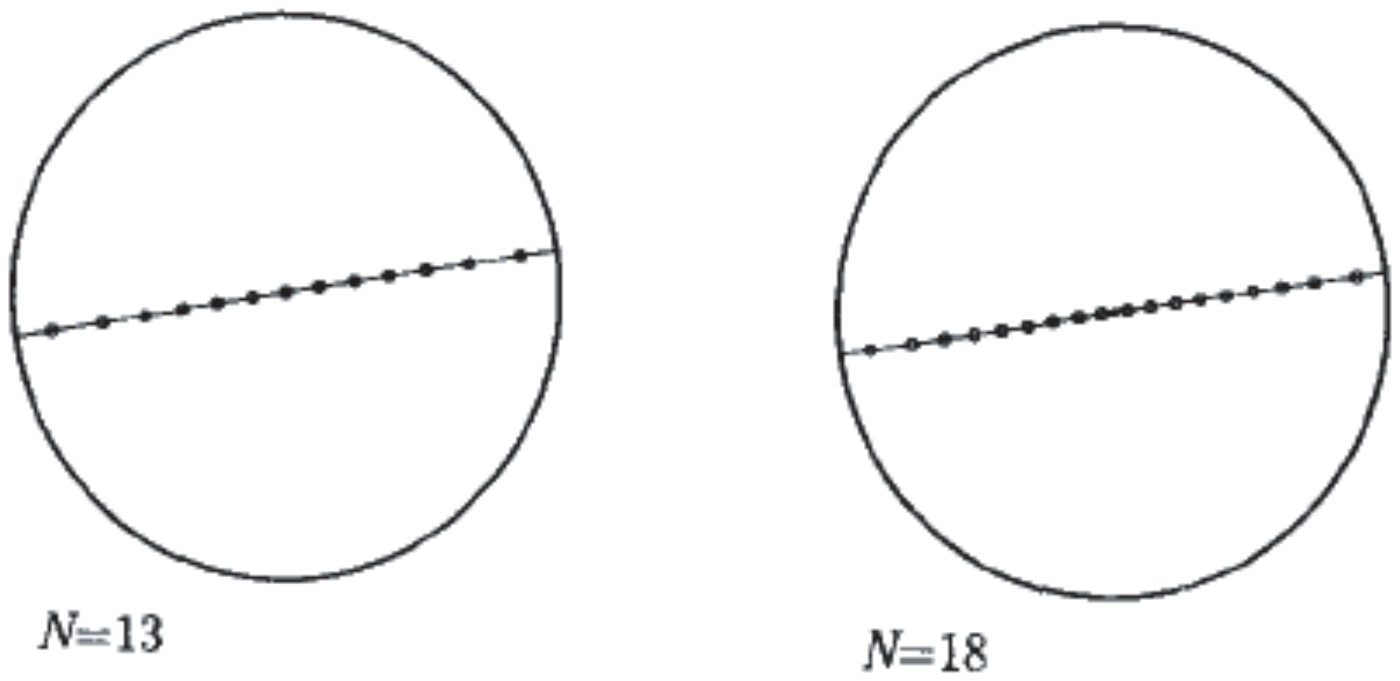}
\caption{
Straight vortex sheet is formed in the interval $12<N<19$. The meron number $N(\Omega) $ in this configuration does not depend on the soliton tension, since its area of the soliton does not change within this interval. Eq.(\ref{PlanarSheetImages}) gives  
$N(\Omega)=0.42 N_0(\Omega)$.
}
 \label{Planar-figure}
\end{figure}


\section{Planar vortex sheet}

For small $\Omega$ the finite size is present:  surface tension of the soliton becomes important, and as a result the meron number is smaller than that which corresponds to the macroscopic solid body rotation of the liquid, $N(\Omega)< N_0(\Omega)$.  Note that the vortex sheet itself performs the solid body rotation together with the container, but the flow of the whole liquid does not follow the solid 
 body rotation. The same deficit of the effective density 
of the liquid involved into rotation has been considered  by  I.M. Lifshitz and M.I. Kaganov  \cite{ILifshitz1955} for the Landau-Lifshitz concentric sheets.

In our case of the vortex sheet connected to the side walls of container, the shape of sheet
depends on angular velocity $\Omega$ and thus on the number $N$ of merons. There is the range of velocities,  where the
equilibrium  configuration of the vortex sheet  is the straight soliton
 along the
diameter of the cylinder (the plane which contains the axis of cylinder). In Fig. \ref{growth-figure}
this is the range $12<N< 19$. For the straight soliton the function $N(\Omega)$ can be found analytically.

Let
  $x$ be the coordinate along the straight soliton with $x=0$ on the axis of
container and $n(x)$ is the density of vortex-merons in the soliton (the
$2\pi$-vortices).  If one neglects the image forces from
the boundary, the equation for $n(x)$ is 
\begin{equation}
  \frac{\hbar}{2m}\int_{-R}^R
dy\frac{n(y)}{x-y}=\Omega x \,.
\label{PlanarSheet}
\end{equation}
This corresponds to the solid
body rotation of the soliton (the soliton is staionary in the rotating frame if superflow velocity produced by other
vortices at the place of a given meron is equal to the solid
body velocity $v_y=\Omega x$). The soliton surface tension does not enter since the
length of the soliton is always the same, $L=2R$.
The equation (\ref{PlanarSheet}) has solution
\begin{equation}
  \frac{\hbar}{2m}   n(x)=\frac{\Omega}{\pi}\sqrt{R^2-x^2} \,.
\label{PlanarSheetSolution}
\end{equation}
The total number of merons is $N=\int dx ~n(x)=N_0/2$.
This should be corrected by 
adding the images of vortices. With vortex images, the Eq.(\ref{PlanarSheet})
transforms to
\begin{equation}
  \frac{\hbar}{2m}\int_{-R}^R
dy~n(y)\left(\frac{1}{x-y}- \frac{1}{x-\frac{R^2}{y}}\right)=\Omega x \,,
\label{PlanarSheetImages}
\end{equation}
which gives $N\approx 0.42 N_0$.

\begin{figure}
\includegraphics[width=1.0\linewidth]{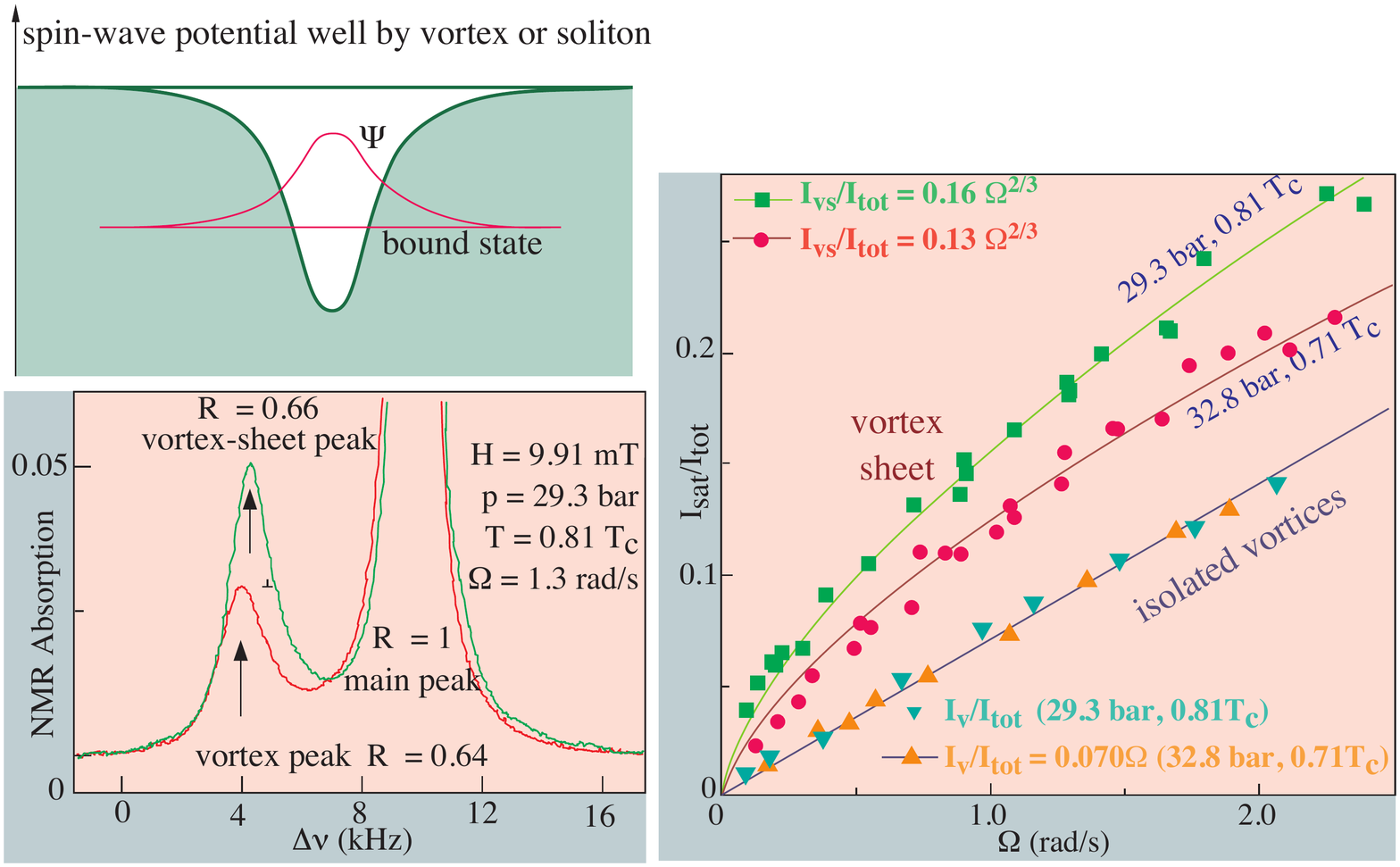}
\caption{
NMR absorption on skyrmions and on the vortex sheet. {\it top left}:  The core 
of skyrmion and the soliton represent correspondingly the 2D and 1D potential wells for spin waves -- magnons. In the NMR experiments  the bound states of magnons in these potentials are excited. 
This corresponds to the satellites in the NMR spectrum on the lower frequency side of the main peak ({\it bottom left}). The vortex sheet state has higher absorption than the lattice of isolated skyrmions, since the area of the soliton is larger than the area of the vortex cores. {\it right}: Intensity of the satellite peak as a function of the angular velocity $\Omega$ of rotation.
In the vortex lattice state the intensity is proportional to the number of vortices, $N_0(\Omega) =(2m/\hbar)\Omega R^2$, and thus is linear in $\Omega$.  In the vortex sheet state 
the intensity is proportional to the area of the soliton, and thus to $\Omega^{2/3}$, which agrees with the Landau-Lifshitz equation.
}
 \label{Satellite-figure}
\end{figure}


\section{NMR signature of the vortex sheet}

The properties of different textures in $^3$He-A (solitons, vortices and vortex sheets) are investigated using NMR, see e.g. review papers \cite{SalomaaVolovik1987,Thuneberg1995}. The cores of these topological oblects produce the potential wells for the spin wave (magnons) in Fig. \ref{Satellite-figure} ({\it top left}).
Excitation of magnons in the bound state gives rise to the satellite peak in the NMR absorption spectrum at the frequency below the main peak. The position of the satellite peak indicates the type of the object, while the intensity of the peak gives the information on the size of the object or on the number of the identical objects.

If the rotating state represents the array of vortex-skyrmions, the intensity is proportional to the number of vortices, and thus is linear in $\Omega$, Fig. \ref{Satellite-figure} ({\it right}). 
In the vortex sheet state of rotation the NMR absorption
in terms of $\Omega$ is close to the $\Omega^{2/3}$ law, Fig. \ref{Satellite-figure} {\it right}. This is consistent with the Landau-Lifshitz equation (\ref{b}).  The area of the
soliton is $A=V/b$, where $V$ is the volume of the cell, and according to the Landau-Lifshitz equation one has
$A\propto\Omega^{2/3}$.

In the vortex-sheet state the measured absorption is 1.5-3 times higher than
in a pure vortex-skyrmion state, Fig. \ref{Satellite-figure} ({\it bottom left}). The vortex sheet produces  the potential well for
magnons, which area per one quantum of circulation   is  larger than
the area of the core of skyrmion.

\begin{figure}
\includegraphics[width=1.1\linewidth]{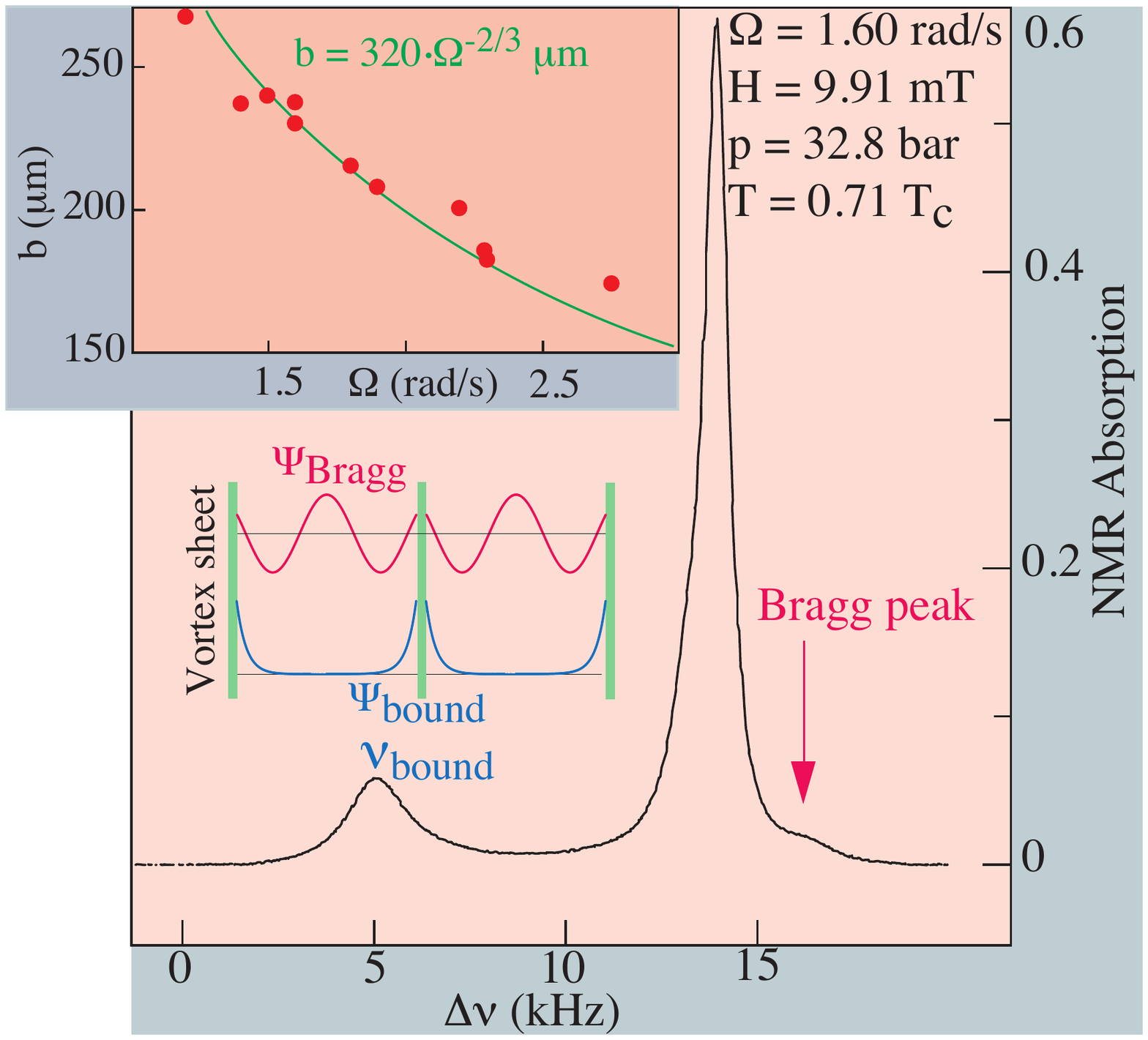}
\caption{
({\it main frame}): There is a small peak in the NMR spectrum on the higher frequency side of the main peak. 
It corresponds to the Bragg reflection satellite from standing spin waves between the neghbouring sheets.
The experimental curve for the distance $b(\Omega)$ between the sheets ({\it top left}) 
 is in agreement with the Landau-Lifshitz equation (\ref{b}) within 10$\%$ accuracy.
}
 \label{Bragg-figure}
\end{figure}


\section{Bragg peak as signature of layered structure 
of vortex sheet}

Locally the folded vortex state represents equidistant
layers, and thus has a local order of
the smectic liquid crystal (Fig. \ref{scales-figure} {\it top left}). 
In NMR experiments such periodic structure is manifested by the Bragg peak in Fig. \ref{Bragg-figure}.
The measured distance $b$ between the sheets in Fig. \ref{Bragg-figure} ({\it top left}) 
 is in a good agreement with the Landau-Lifshitz equation (\ref{b}), where the known values of the superfluid density and the soliton tension are used.

\begin{figure}
\includegraphics[width=1.2\linewidth]{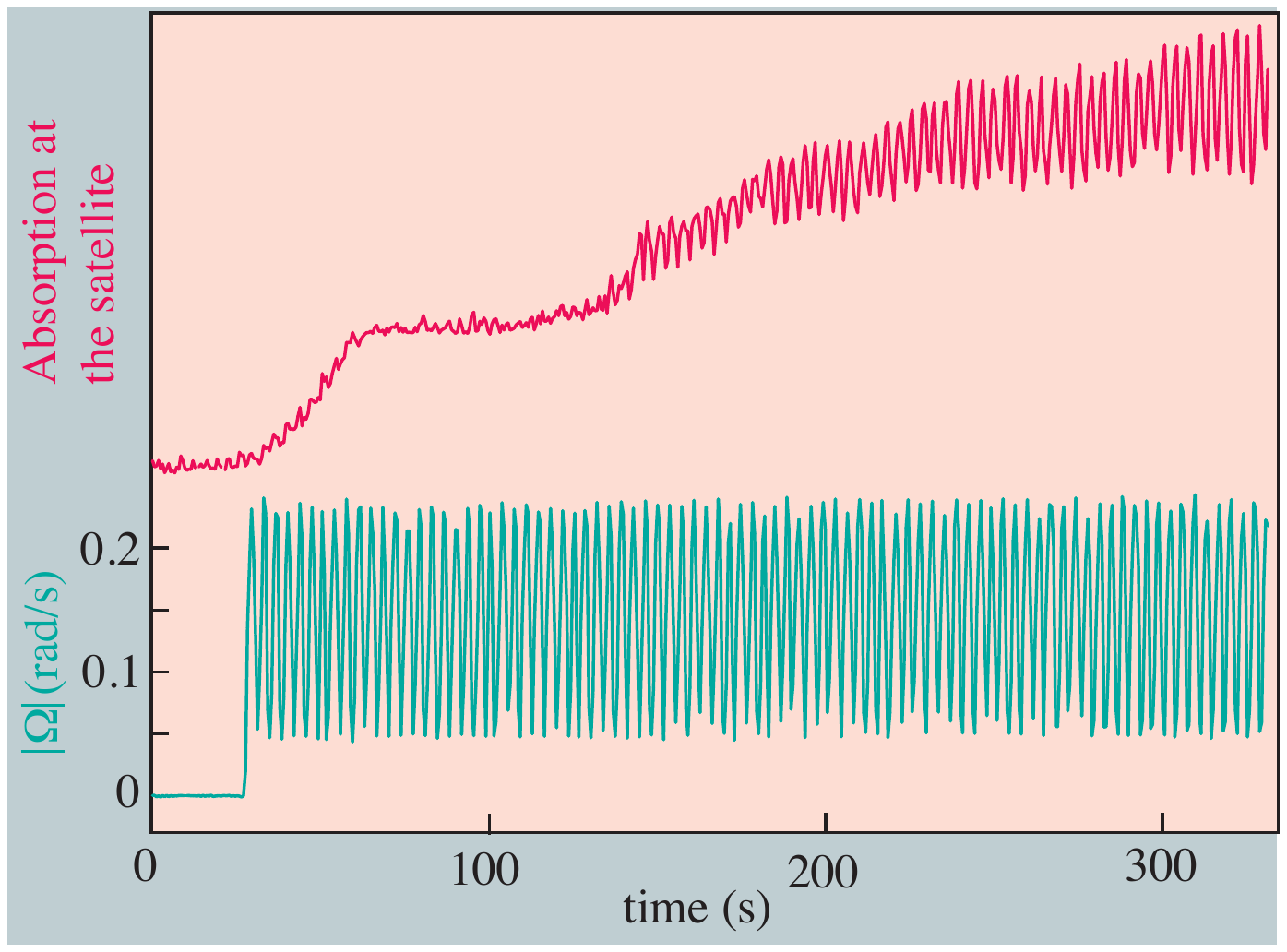}
\caption{
Straight soliton is formed by periodic oscillation of container. The formation of this soliton is indicated by plateau in the absorption, since the area of the straight soliton does not depend on the number of merons. After the straight soliton is formed, the folded vortex sheet is obtained by adiabatic growth of the soliton by gradual increasing of the rotation velocity. }
 \label{SolitonFormation-figure}
\end{figure}

\section{Formation of the vortex sheet. Experiment.}
\label{NucleationSection}

 In experiment, the vortex sheet  is
reproducibly created  during the sinusoidal modulation of velocity  
$\Omega =  \Omega_0 \sin(\omega t)$, see Fig. \ref{SolitonFormation-figure} {\it bottom}.
The
period of modulation is usually $T=2\pi/\omega \approx 10$~s. The vortex sheet appears if 
the amplitude $\Omega_0$ of modulation exceeds  some critical velocity which is 
slightly above the critical velocity for the nucleation of the skyrmions. This 
 means that the formation of the
vortex sheet is somehow connected to the presence of skyrmions in the cell.
During multiple and
fast alteration of the angular velocity of the container  the skyrmions
cannot maintain the equilibrium arrangement corresponding to a
momentary angular velocity. They are  pressed to the side wall of
the container, where their mutual collapse supposedly gives rise to a small seed
of the soliton attached to the surface of container.  

Anyway, once the soliton
is created, the critical velocity for entering 
new vorticity -- merons -- into the  soliton from the wall is found to be essentially smaller   than the critical velocity  for the creation of isolated skyrmions.  Thus the soliton is able to grow by absorbing the
merons created at the wall. Finally the soliton with accumulated merons forms the folded
vortex sheet connected with the side wall.

   The upper frame of Fig. \ref{SolitonFormation-figure} displays the absorption at the satellite  as a function of time. In the beginning of the modulation the 
increase of the absorption is small and corresponds to nucleation of few 
separate continuous vortices. After 2.5 minutes the first seed of the 
vortex-soliton is created and devolops within $\approx$ 20 seconds to 
constant level, where there is no sinusoidal structure in absorption. This
 corresponds to the formation of the straight soliton in Fig. \ref{Planar-figure}. 
During periodic drive the absorption remains constant, because the area of 
the vortex sheet does not change.  The intensity of the
satellite also agrees with the straight soliton. The
numerical simulation shows that the straight shape of the vortex
sheet is the stable configuration in the certain window of the
meron number $N$ in the soliton:  $12 <N < 19$
\cite{Heinila1995}.  

When $\Omega$ is increased and exceeds some critical value
the straight soliton becomes unstable against the
curved one. At large  $N$ and  $\Omega$ the length of the soliton grows and 
finally the folded vortex sheet sweeps all the cell to produce the homogeneous distribution
of merons, which in average imitates the solid body rotation of superfluid,
$<{\bf v}_{\rm s}>={\bf \Omega}\times {\bf r}$. Instead of one scale, which
characterizes the intervortex distance in the conventional vortex array, the
vortex sheet is characterized by two scales: intervortex distance within
the sheet and the mean distance $b$ between the neighbouring parts of the curved
soliton.

\begin{figure}
\includegraphics[width=1.0\linewidth]{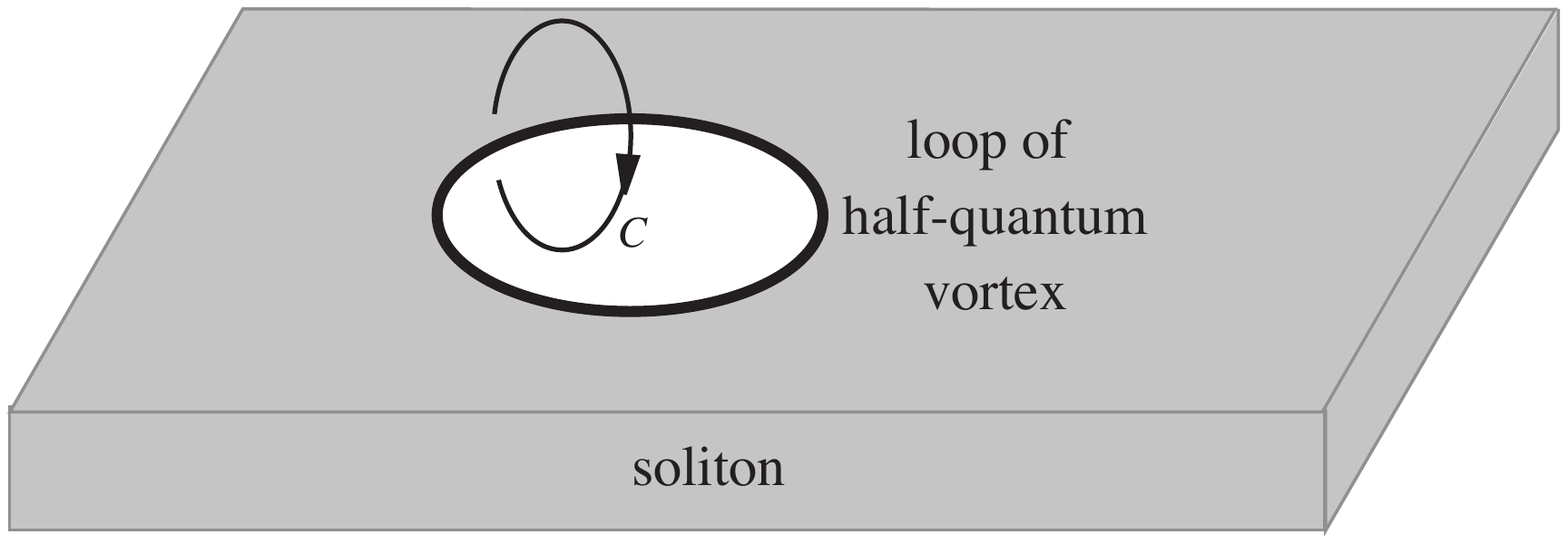}
\caption{
 Stability of the vortex sheet. Vortex sheet is based on the topological soliton. As distinct from the  ferromagnetic domain walls, it is possible to drill a hole in the soliton. Such hole is bounded by the
topological line defect -- the string loop, where the string is the half-quantum vortex (vortex with winding number $N=1/2$). If the vortex loop with the radius exceeding the critical value on the order of the 
thickness of the soliton wall ($\sim 10~\mu$m) is created, the loop will further grow and destroy the whole vortex sheet. It is known that neutron irradiation produces the hot spots 
(micro Big-Bang \cite{NeutronIrradExp}) of the proper size, but all the attempts to destroy the vortex sheet by the neutron irradiation failed. 
}
 \label{Hole-figure}
\end{figure}


\section{Topological stability of the vortex-sheet }

   The topological stability of both the
soliton and the kinks )merons) inside the soliton prevents the vortex sheet
from breaking into the individual vortices and thus the sheet is
extended from one boundary to another. The termination of the vortex
sheet in the bulk can occur only if the termination line represents
the half-quantum vortex \cite{VolovikMineev1976}. This is the object
with the singular core of the coherence length size (hard core), which is the
counterpart of the Alice string in high energy physics
 and of the object with the one-half of the
magnetic flux quanta observed in high temperature supercoductors
\cite{Volovik2003}. 

 If one manages to create the loop of the half-quantum vortex within the soliton 
 (Fig. \ref{Hole-figure}) with the radius exceeding the critical value on the order of the soliton thickness 
 ($\sim 10~\mu$m), this vortex ring  will grow and destroy the whole vortex sheet.
It is known that in
superfluid $^3$He-B  the hard-core vortices of similar size are produced by neutron irradiation
 \cite{NeutronIrradExp}. 
So we tried to make a hole in the vortex sheet in a similar way.  However, irradiation by neutrons and by $\gamma$-quanta for several hours could not destroy the vortex sheet.

\begin{figure}
\includegraphics[width=1.0\linewidth]{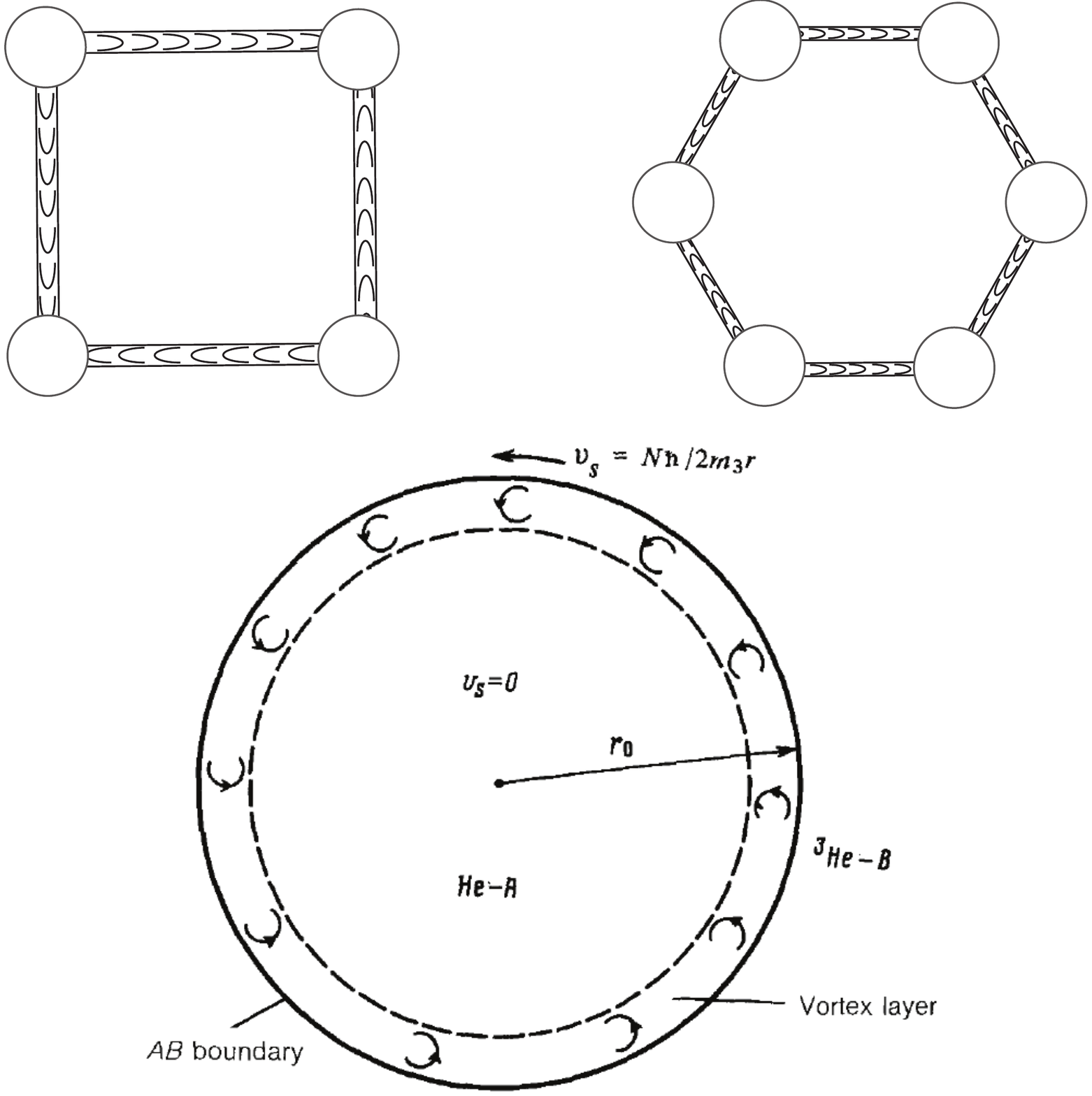}
\caption{
Multiple vortices. Closed vortex sheets represent $N$-quantum vortices. ({\it top}): Vortex sheets with $N=4$ and $N=6$ merons in $^3$He-A.
The radius of the $N$-quantum vortex is determined by the balance of forces: the repulsive interaction between the vortex-merons is compensated by the attraction due to surface tension of the soliton,
see Eq.(\ref{MultipleRadiusPoly}).
 ({\it bottom}): Multiply quantized vortices in $^3$He-B. Vorticity in the form of the vortex sheet can be concentrated 
at the interface between $^3$He-B and the cylindrical domain of $^3$He-A \cite{VolovikMisirpashaev1990}, see also Sec. \ref{Sec:AB} and  Fig. \ref{ABsheet-figure} {\it right bottom}. 
}
 \label{Hexagon-figure}
\end{figure}

\section{Cylindrical vortex sheets 
as multiply quantized vortices}

There are many possible equilibrium configurations of the rotating chiral superfluid $^3$He-A, which obey the solid body rotation of the liquid on a macroscopic scale. Practically all of them (including the array of vortex-skyrmions) are metastable. But if they are somehow created, they live practically forever. 
The main problem is to find the proper way for the preparation of particular configurations.
There are many topological objects, which still avoided observation, because the scenarios for their creation have not been found. Among them are the half-quantum vortex and multiply quantized vortices.

The vortex sheet provides the possibility for the construction of multiply quantized vortices 
\cite{VolovikKrusius1994}.
The closed vortex sheet with $N$ merons represents the $N$-quantum vortex. Since the $N=2$ vortex is the known vortex skyrmion, and the odd number of merons in the closed vortex sheet is prohibited by topology, the  multiply quantized vortices start with $N=4$, see Fig. \ref{Hexagon-figure}.

 For large $N\gg 1$, the multiply quantized vortex is the cylindrical 
vortex sheet. Its radius $r_N$ is obtained by minimization of the superflow energy outside the
vortex $+$ the energy of the soliton wall:
\begin{equation}
 E_N=\pi \rho_{\rm s} \frac{\hbar^2}{4m^2}N^2~  {\rm ln} \frac{R}{r_N}+
2\pi \sigma r_N\,.
\label{multiple}
\end{equation}
Here $R$ is the radius of the cell.
Minimization gives 
\begin{equation}
 r_N=N^2 \frac{\hbar^2 \rho_{\rm s}}{8m^2\sigma}\,.
\label{MultipleRadius}
\end{equation}
 For small $N$ (but for $N>3$) the cylinder should be substituted by the polyhedron -- the $N$-gonal prism, Fig. \ref{Hexagon-figure}. The radius of the prism is determined by the balance of forces: the repulsive interaction between vortices is compensated by the attraction due to surface tension of the soliton:
\begin{equation}
 r_N=\frac{\pi(N-1)}{\sin(\pi/N)} \frac{\hbar^2 \rho_{\rm s}}{8m^2\sigma}\,.
\label{MultipleRadiusPoly}
\end{equation}
 Within the order of magnitude, Eqs.(\ref{MultipleRadius}) and (\ref{MultipleRadiusPoly}) give also the correct estimate for the core size of the skyrmion
 in Fig. \ref{Skyrmion-figure}, which consists of $N=2$ merons.

\begin{figure}
\includegraphics[width=1.2\linewidth]{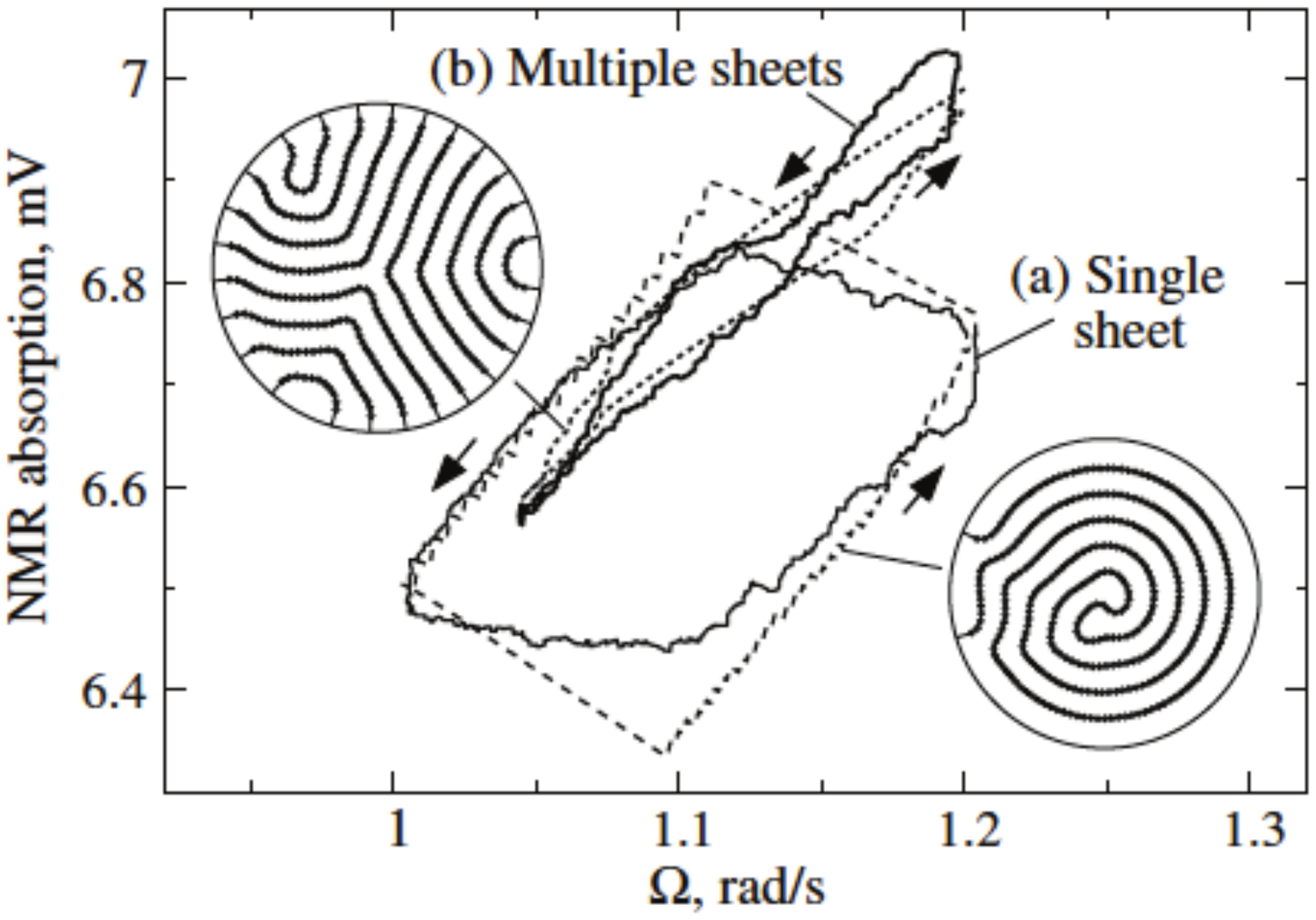}
\caption{
Selection of rotating states by fast dynamics. {\it center}: The hysteresis loop in the state with the
multiple vortex sheets is narrow compared to that  in the state with a single folded sheet. In other words, the multiple sheets 
have much faster dynamics than the 
folded sheet. That is why the multiple sheets are better adjusted to the rotation, if the rotation is accompanied by rapid oscillations.
 {\it left insert}: numerical simulations for the configuration with multiple sheet.
 {\it right insert}: numerical simulations for the configuration with adiabatically grown folded sheet.
}
 \label{MultipleSheet-figure}
\end{figure}


\section{Multiple sheets}
\label{Sec:MutipleSheets}

Though the array of small cylindrical sheets is energetically more favorable than the folded vortex sheet, we failed to produce them. Probably for that it is necessary to construct the proper disturbance on the container wall, which makes the creation of the cylindrical walls easier than the creation of skyrmions
and of the vortex sheets connected to the wall. 

Instead the way was found of how to split the folded vortex sheet into several pieces, which are connected to the side wall of container. The splitting occurs when the original single sheet is subjected to a high-frequency modulation with large amplitude \cite{Kopnin2002}, see Fig. \ref{MultipleSheet-figure}.  The obtained structure is determined by dynamics of merons. The narrow hysteresis loop in Fig. \ref{MultipleSheet-figure} corresponds to multiple sheets, which are formed after modulation  $\Omega(t)= (1.2+0.4 \sin (\omega t))$ rad/s, with $2\pi/\omega= $10 s.
The response of the short sheets is faster than that of the folded sheet, since with the radial alignment of the sheets it is easier for merons to enter and escape the soliton during the fast modulation of angular velocity. 

\begin{figure}
\includegraphics[width=1.1\linewidth]{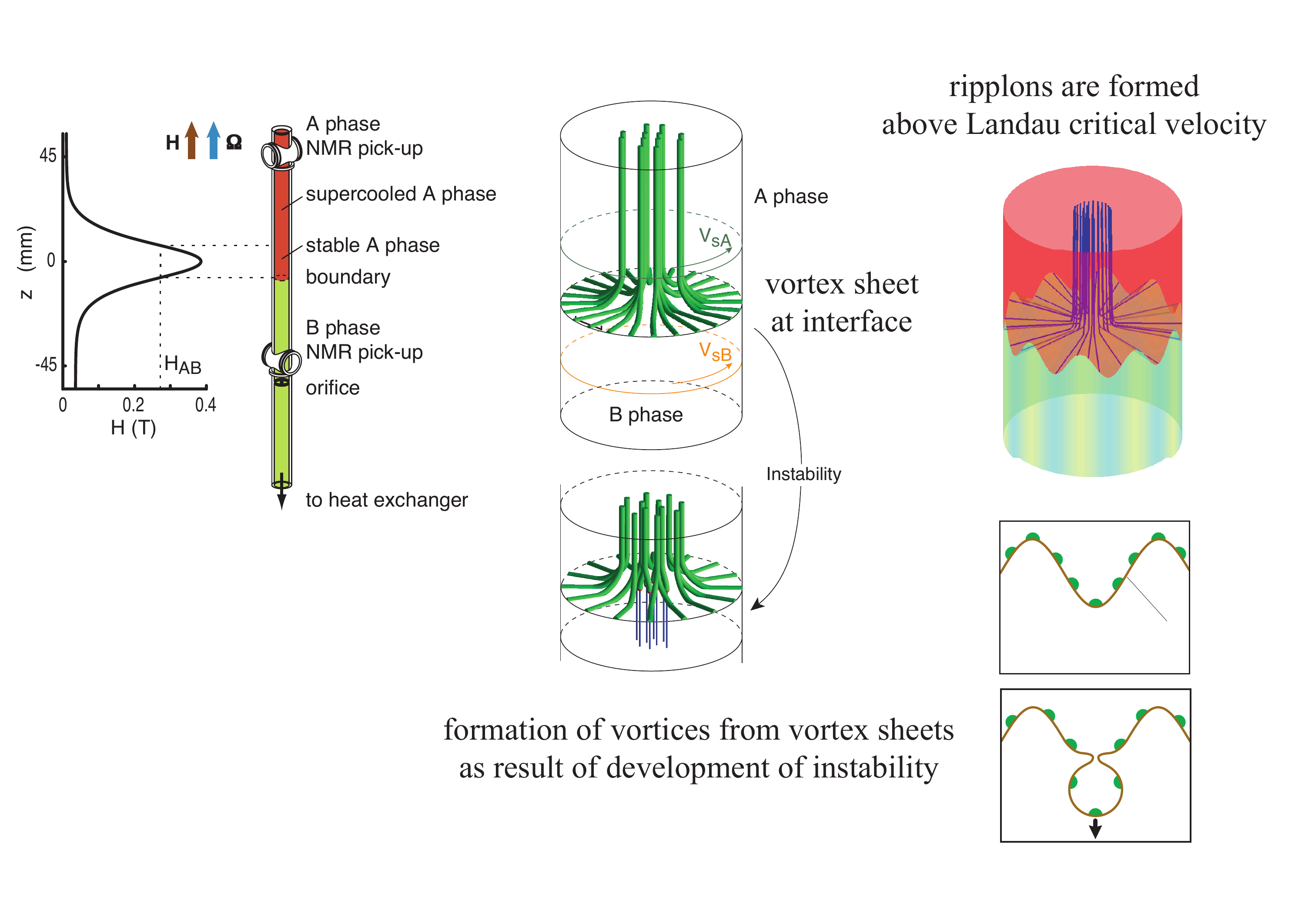}
\caption{
Vortex sheet at the interface between rotating $^3$He-A and the vortex-free $^3$He-B.
({\it left}): Experimental cell. The enhanced magnetic field at $z=0$ prohibits propagation of the B-phase 
into the upper part of the cell. ({\it center top}): Vortex-full A-phase is separated from the vortex-free B-phase by the vortex sheet, which is concentrated at the AB interface. That is why the interface serves 
as tangential velocity discontinuity. ({\it right top}):
When the velocity jump across the interface reaches the critical value, ripples are formed (analogs of the wind-generated surface waves). 
({\it right bottom}):  The development of the instability leads to the formation of the droplet of the vortex sheet inside the B-phase. Such sheet is unstable to formation of the quantized B-phase vortices ({\it center bottom}).
}
 \label{ABsheet-figure}
\end{figure}


\section{Vortex sheet at AB interface}
\label{Sec:AB}

A crucial property of $^3$He superfluids is the existence of the several
length scales. In particular, the core size of a skyrmion in $^3$He-A
is $10^3$ times larger than the core size of quantized vortices in  $^3$He-B.
The sharp (hard-core) vortices are not easily created: the critical velocity of the vortex nucleation under rotation is inversely proportional to the core radius \cite{Parts1995}. As a result one can prepare the state in the rotating cryostat, in which the A-phase skyrmion lattice simulates the macroscopic solid body rotation of superfluid,
while  the B-phase is still in  the static vortex-free state, which is called the Landau state. 
The A-phase vortex-skyrmions cannot penetrate from the A- to
the B-phase, so they form a surface vortex sheet at the phase boundary \cite{Hanninen2003}, Fig. \ref{ABsheet-figure} {\it center top}. Vortex sheets, which appear in the mixture of several superfluids,
have been discussed in Ref. \cite{VolovikMisirpashaev1990} (see the cylindrical vortex sheet  
in Fig. \ref{Hexagon-figure} {\it bottom}), and in Ref. \cite{KasamatsuTsubota2009}.  

The vortex sheet in Fig. \ref{ABsheet-figure} {\it center top} separates two superfluids moving with respect to each other.
Near the boundary of the container the jump in the tangential velocity reaches $|v_{\rm A}-v_{\rm B}|= \Omega R$. This arrangement brings a lot of interesting physics. 

If the velocity difference $|v_{\rm A}-v_{\rm B}|$ exceeds the critical value, the interface experiences the instability towards the generation of the surface waves -- ripplons (analogs of the capillary-gravitational waves in conventional lquids),  Fig. \ref{ABsheet-figure} {\it top right}. The onset of the analog of the Kelvin-Helmholtz instability is marked by the appearance of the vortex lines in $^3$He-B which are detected in NMR measurements \cite{Blaauwgeers2002}. 
At the non-linear stage of the instability the droplet of the vortex sheet is formed, Fig. \ref{ABsheet-figure} {\it right bottom} and  Fig. \ref{Hexagon-figure} {\it bottom}). It propagates into  the B-phase, where it splits into singly quantized vortices, which form the vortex cluster in Fig. \ref{ABsheet-figure} {\it center bottom}. Formation of vortices in the B-phase after instablity allowed us to measure in the NMR experiments the value of the critical velocity and compare it with the theory.

 It appeared that the measured threshold of the instability is lower than that for the classical Kelvin-Helmholtz instability, and instead it satisfies the Landau critical velocity for the radiation of ripplons, see details in \cite{Volovik2002} and in Chapter 27 of the book \cite{Volovik2003}. In the shallow water limit this corresponds to the Zel'dovich--Starobinsky effect of radiation of the electromagnetic waves by rotating black holes
 (Chapter 31 in Ref. \cite{Volovik2003}).


\section{From vortex sheet to quantum turbulence}

\begin{figure}
\includegraphics[width=1.1\linewidth]{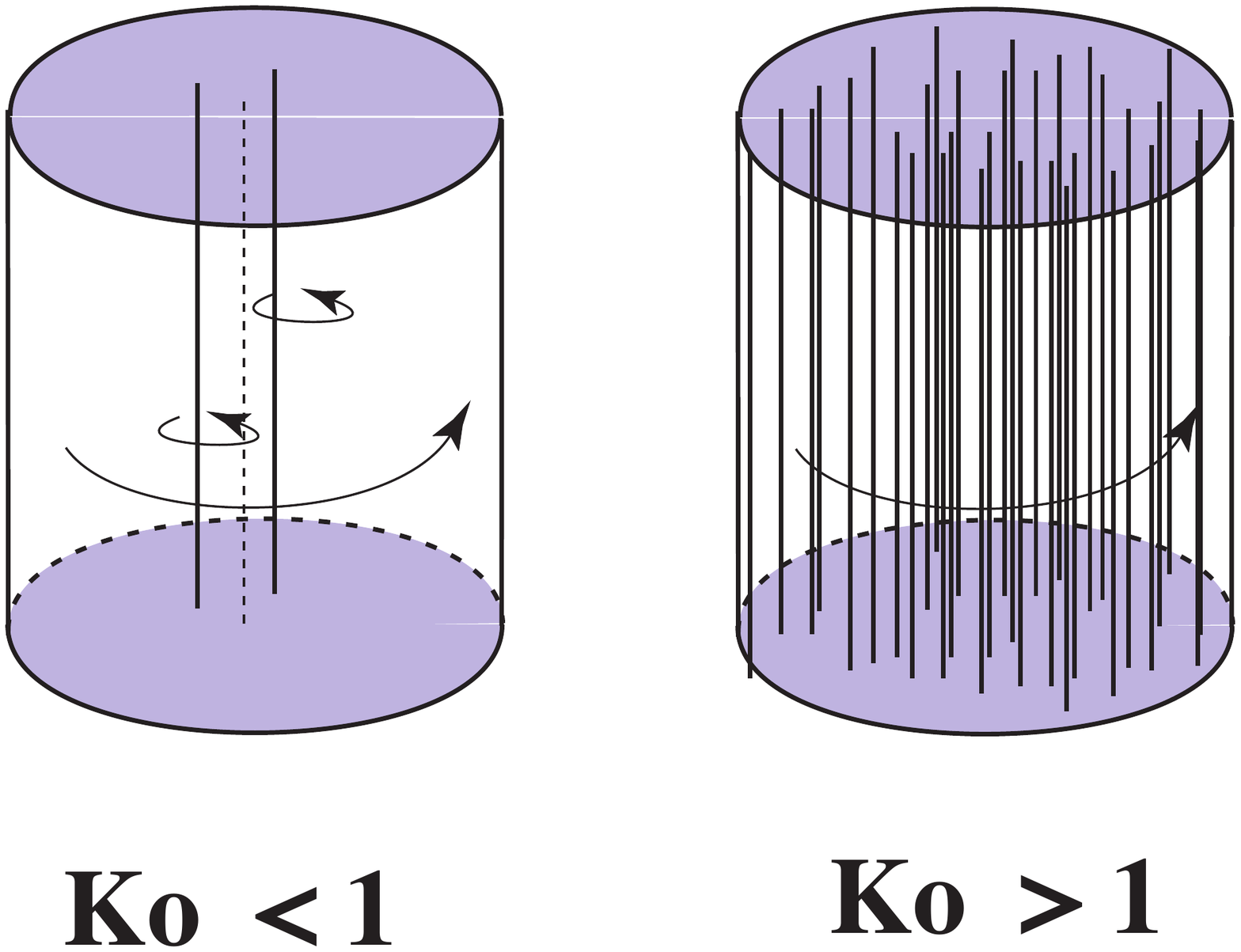}
\caption{Results of the injection of few vortices into $^3$He-B after development of the vortex sheet instability at AB interface in Fig. \ref{ABsheet-figure} ({\it right}). The final state of $^3$He-B under rotation depends on the value of Kopnin number ${\rm Ko}(T)$ --  the temperature dependent  analog of the Reynolds number, which characterizes the quantum turbulence. As distinct from the Reynolds number in classical liquids, the Kopnin number does not contain velocity of the liquid.
For ${\rm Ko}(T) < 1$, the injected vortices form a small cluster in the center of container. For ${\rm Ko}(T) > 1$, the injected vortices are multiplied forming the chaotic vortex flow, which fills the whole cell.  Finally the turbulent vorticity relaxes to the equilibrium lattice of the rectilinear vortices, which imitates the solid body rotation of the B-phase.}
 \label{KoNumber-figure}
\end{figure}

Formation of the B-pase vortices due to the instability of the surface vortex sheet allows us to use this phenomenon as the working tool for injection of vortices into the vortex-free B-phase under rotation.
The injection of few vortices into superfluid in its metastable vortex-free Landau state revealed the new phenomenon in quantum turbulence \cite{Finne2003}.
The NMR measurements after injection demonstrated the sharp transition to turbulence. At temperature above $0.60\,T_c $ (where $T_c $ is the transition temperature for superfluidity)  the flow of superfluid is regular (laminar): the injected vortices from a small cluster in the center of the cell, Fig. \ref{KoNumber-figure} ({\it left}) and Fig. \ref{ABsheet-figure} ({\it center bottom}).  However, below this temperature the turbulent behaviour is observed: vortices are multiplied in the turbulent regime and finally fill the whole region of the B-phase, Fig. \ref{KoNumber-figure} ({\it right}). 

Suprisingly, the observed sharp transition to turbulence is insensitive to the fluid velocity. This is in a striking contrast to classical turbulence regulated by the Reynolds number ${\rm Re}=UR/\nu$, which is proportional to the velocity $U$ of the liquid. The quantum superfluid turbulence is controlled by an intrinsic parameter of the superfluid: the Kopnin number ${\rm Ko}$. It is the ratio of the reactive and dissipative parameters in the equations for superfluid hydrodynamics, which depends only on temperature $T$. 

In the Fermi superfluids, such as superfluid $^3$He and superconductors, the Kopnin number ${\rm Ko} \sim \omega_0 \tau$, where $\omega_0$ is the minigap -- the distance between the levels of Andreev-Majorana fermions in the core of the vortex --  and $\tau$ is their lifetime. For ${\rm Ko}(T) < 1$ the effect of chiral anomaly (spectral flow) provides the momentum exchange between vortices and the normal component of the liquid. In this regime dissipation is dominating, and the flow of the superfluid component is laminar.
For ${\rm Ko}(T) > 1$  the effect of the inertial term is dominating, and the laminar flow becomes unstable towards turbulence.

 \section{Conclusion}

The vortex sheet structure of rotating superfluid has been calculated by Landau and Lifshitz in 1955. In 1995, i.e. 40 years later, it was experimentally demonstrated that for the complicated superfluids with the multi-component order parameter this concept was instrumental. The Landau-Lifshitz  vortex sheet proved to be an important physical object with nontrivial topology and with many physical applications.

\section*{Acknowledgements}
This work has been supported in part  by the Academy of Finland
(project no. 250280), and by the facilities of the Cryohall
infrastructure of Aalto University.

\end{document}